\documentclass[preprint]{aastex}

\slugcomment{Accepted for publication in
\it The Astrophysical Journal, Letters}

\shorttitle{SN~1991T and the Value of H$_0$}
\shortauthors{Gibson \& Stetson}

\begin{document}

%USER-DEFINED DEFINITIONS
\def\kms{km\,s$^{-1}$}
\def\spose#1{\hbox to 0pt{#1\hss}}
\def\simlt{\mathrel{\spose{\lower 3pt\hbox{$\mathchar"218$}}
     \raise 2.0pt\hbox{$\mathchar"13C$}}}
\def\simgt{\mathrel{\spose{\lower 3pt\hbox{$\mathchar"218$}}
	  \raise 2.0pt\hbox{$\mathchar"13E$}}}
\def\eg{{\rm e.g.}}
\def\ie{{\rm i.e.}}
\def\etal{{\rm et~al.}}

\title{Supernova 1991T and the Value of the Hubble Constant}

\author{Brad K. Gibson}
\affil{Centre for Astrophysics \& Supercomputing, 
Swinburne University,\\
Mail \#31, P.O. Box 218, Hawthorn, Victoria, Australia 3122}
\and
\author{Peter B. Stetson}
\affil{Dominion Astrophysical Observatory, Herzberg Institute of 
Astrophysics,\\\
National Research Council, 5071 W. Saanich Rd., Victoria, B.C., Canada
V9E~2E7}

\begin{abstract}

Based upon multi-epoch \it Hubble Space Telescope \rm observations, we
present the discovery of sixteen high-quality Cepheid candidates in
NGC~4527.  Corrected for metallicity effects in the Cepheid
period-luminosity relation, we derive a distance, including both random
(r) and systematic (s) uncertainties, of 13.0$\pm$0.5(r)$\pm$1.2(s)\,Mpc.
Our result is then used to provide a calibration of the peak $B$-, $V$-,
and $I$-band luminosities of the peculiar Type~Ia supernova SN~1991T, a
resident of NGC~4527.  Despite its documented spectroscopic peculiarities,
after correction for the decline rate--luminosity correlation the
corrected peak luminosity is indistinguishable from those of so-called
``normal'' Type~Ia SNe.  With now nine local calibrators at our disposal,
we determine a robust value for the Hubble Constant of
H$_0$=73$\pm$2(r)$\pm$7(s)\,km\,s$^{-1}$\,Mpc$^{-1}$.  \end{abstract}


\section{Introduction}
\label{introduction}

When corrected for the shape of their light curves, the peak apparent
brightnesses of Type~Ia supernovae (SNe) provide an outstanding secondary
distance indicator (\eg, Phillips et~al. 1999; Jha et~al. 1999; Saha
et~al. 1999; Gibson et~al. 2000a).  First, the extreme luminosity of the
SNe allows one to probe substantially further into the unperturbed Hubble
flow, in comparison with competing secondary indicators such as the
Tully-Fisher relation and surface brightness fluctuations.  Second, the
intrinsic scatter is smaller than that of other indicators.  While
attempts have been made to provide a purely theoretical calibration for
the peak luminosity (\eg, H\"oflich \& Khokhlov 1996), remaining model
uncertainties still favor the observational approach.  The \it
Sandage/Tammann/Saha 
Type~Ia SN HST Calibration Project \rm (Saha et~al. 1999, and
references therein) was designed to provide just such a calibration, by
determining the Cepheid distances to eight nearby galaxies (IC~4182,
NGC~5253, 4536, 4496A, 4639, 3627, 3982, 4527) host to Type~Ia SNe.  This
original program has been supplemented with two additional calibrators
situated within NGC~3368 (Tanvir et~al. 1999) and NGC~4414 (Turner et~al.
1998).  Table~1 of Gibson et~al. (2000a) provides a complete list of the
galaxies and SNe covered by all these calibration programs, including a
subjective quality ranking.

Of the ten nearby Type~Ia SNe targeted by the various teams, only one is
classified as spectroscopically peculiar --- SN~1991T.  While the data
suggest that SN~1991T was a supernova with peculiar surface abundances (no
evidence for Si, Ca, or S absorption during the early phases of the
explosion), and had a light curve in $B$ and $V$ that was marginally
broader than usual, in most other ways it seems to have been a normal
Type~Ia explosion (Phillips et~al. 1992).  Derivation of the Cepheid
distance to NGC~4527, the host galaxy to SN~1991T, provides a direct test
of the robustness of the technique as applied to potentially peculiar SNe.

In what follows, we present the Cepheid distance to NGC~4527, provide a
calibration for the peak $B$-, $V$-, and $I$-band luminosities of
SN~1991T, and present an improved determination of the Hubble Constant. As
will be seen, the derived peak luminosities do not appear to be affected
by the spectroscopic peculiarities of the SN.

\section{Analysis}
\label{analysis}

The SAB(s)bc spiral galaxy NGC~4527 was observed at 17 epochs over the
69\,d window spanning 04/11/99-06/19/99, as part of the \it 
Sandage/Tammann/Saha
Type~Ia Supernova Calibration Project \rm (\it HST \rm PID\#7504;  Saha
et~al. 1999, and references therein). As for previous galaxies in this
program (and the \it HST Key Project on the Extragalactic Distance
Scale\rm), cosmic-ray split 2500\,s exposures in F555W ($V$: 12 epochs)
and F814W ($I$: 5 epochs) were taken.

Following the well-documented methodology of the \it HST Key Project \rm
(\eg, Gibson et~al. 2000a, and references therein), the data were
processed with ALLFRAME (instrumental photometry -- Stetson 1994) and
TRIAL (calibration and variable finding -- Stetson 1996).\footnote{An
independent reduction will be provided by the Sandage/Tammann/Saha 
Team (Saha
et~al. 2001), but at the time this paper was submitted their results were
not yet available.} As was done in Gibson et~al. (2000a) and Freedman
et~al.  (2001), the WFPC2 photometric zero point and charge-transfer
corrections adopted are essentially those of Stetson (1998), which are
nearly identical to those of Whitmore et~al. (1999) and Dolphin (2000).
The quoted systematic uncertainty in our final result incorporates a
component which allows for imprecision in our current understanding of the
spatial and temporal variations of the WFPC charge-transfer inefficiency.
An initial candidate list of 25 Cepheids was identified by TRIAL;
light-curve inspection reduced this number to a final list of 16 high
quality Cepheids, the properties of which are listed in Table~1.
Epoch-by-epoch photometry for each of the Cepheids, local calibration
standards, and accompanying light curves have been made available on the
HST Key Project archive.\footnote{{\tt 
http://www.swin.edu.au/astronomy/bgibson/H0kp/} and \hfil\break
{\tt http://www.ipac.caltech.edu/H0kp/}}

In Figure~1 we show the apparent $V$- and $I$-band Period-Luminosity (PL)
relations for the 16 Cepheids in NGC~4527 (upper and middle panels,
respectively).  The apparent moduli, in conjunction with (i) the
assumption of a standard reddening law, (ii) an LMC true modulus
$\mu_\circ$(LMC)=18.45$\pm$0.10\,mag (Freedman et~al. 2001), and (iii) a
set of LMC PL relations (Udalski et~al. 1999), leads to the distribution
of de-reddened moduli shown in the lower panel, the mean of which implies
a true modulus $\mu_\circ$=30.482$\pm$0.085\,mag
($d_\circ$=12.5$\pm$0.5\,Mpc).  The quoted random uncertainty includes
those due to photometry, extinction, and the dereddened PL fit (corresponding
to $R_{\rm PL}$ in Table~7 of Gibson et~al. 2000a).
This distance assumes the Cepheid PL
relation has no metallicity dependency (\ie, $\gamma_{\rm
VI}$=$+$0.0\,mag\,dex$^{-1}$; cf. Kennicutt et al. 1998).

The mean reddening inferred from these Cepheids is
E(V$-$I)=0.268$\pm$0.031; since no Cepheid field \ion{H}{2} region
abundance analysis exists for NGC~4527, it is tempting to infer a
metallicity based upon this line-of-sight reddening.  However, plotting
\ion{H}{2} region metallicity 12$+$$\log$(O/H) versus line-of-sight of
reddening E(V$-$I), for each of the Cepheid fields of the 30 galaxies in
Table~5 of Freedman et~al. (2001), yields what is effectively a scatter
plot over the limits 0.1$\simlt$E(V$-$I)$\simlt$0.3 and
8.50$\simlt$12$+$$\log$(O/H)$\simlt$9.35.  For the derived NGC~4527
Cepheid reddening, 12$+$$\log$(O/H) could plausibly lie anywhere within
this range.  Therefore we adopt the very conservative value of
12$+$$\log$(O/H)=8.9$\pm$0.4 for the metallicity of the Cepheid field in
NGC~4527.  Under this assumption, the true modulus for NGC~4527 increases
by 0.08\,mag for $\gamma_{\rm VI}$=$-$0.2\,mag\,dex$^{-1}$, to
$\mu_\circ$=30.562$\pm$0.085\,mag ($d_\circ$=13.0$\pm$0.5\,Mpc) and will
be used in what follows.

The functional form for the adopted Hubble relations in our analysis 
is described in
Freedman et~al. (2001) and is written
\begin{equation}
{\rm B,V,I}_{\rm max} - 5\log({\rm cz}_{\rm CMB}) =
a(\Delta m_{15}({\rm B})_t-1.1) + b,
\label{eq:eq1}
\end{equation}
\noindent
where, from Phillips et~al. (1999), $\Delta m_{15}$(B)$_t$=$\Delta
m_{15}$(B)$_{\rm obs}$$+$0.1\,[E(B$-$V)$_{\rm Gal}+$E(B$-$V)$_{\rm
Host}$].  E(B$-$V)$_{\rm Gal}$ and E(B$-$V)$_{\rm Host}$ represent the
Milky Way foreground and host galaxy reddenings, respectively, and are
tabulated by Phillips et~al. (1999) for each SN employed in our
analysis.\footnote{For SN~1991T, the adopted foreground$+$host reddening
was E(B$-$V)=0.16$\pm$0.05 (Phillips et~al. 1999; Tbl~2).  The host
reddening was derived from the favored late-time color excess technique;
as noted by Phillips et~al., the color excess at maximum light should not
be employed for 1991T-like events.  Despite the uncertainty in deriving
reddenings for 1991T-like events, only a radical alteration of the
inferred ($\simgt$100\% increase in E(B$-$V), from 0.16 to $\simgt$0.35)
could have more than a 1$\sigma$ effect on the final weighted mean for
H$_0$.} The relevant coefficients and dispersions employed are
$\sigma$(B)=0.15, $a$(B)=$+$1.102$\pm$0.187, $b$(B)=$-$3.677$\pm$0.050,
$\sigma$(V)=0.14, $a$(V)=$+$1.016$\pm$0.160, $b$(V)=$-$3.624$\pm$0.044,
$\sigma$(I)=0.16, $a$(I)=$+$0.975$\pm$0.168, and
$b$(I)=$-$3.259$\pm$0.041, and were derived from the subset of 36 SNe from
the Cal\'an/Tololo (Hamuy et~al. 1996) and CfA (Riess et~al. 1998) surveys
with (i) 3.5$<$$\log$(cz$_{\rm CMB}$)$<$4.5, and (ii)  $|$B$_{\rm
max}$$-$V$_{\rm max}$$|$$<$0.20. Other functional forms (\eg, Phillips
et~al. 1999; Jha et~al. 1999; Gibson et~al. 2000a) lead to results
indistinguishable from those described here. Equation~\ref{eq:eq1},
coupled with the zero points provided by the Cepheid distances to
NGC~4527, 4639, 4536, 3627, 3368, 5253, 4414, 4496A, and IC~4182 provides
a value for the Hubble Constant.

Before deriving H$_0$, we first ensured that the Cepheid distances for the
other eight calibrators were on the same footing as that for NGC~4527, as
described here.  Using the same WFPC2 photometric zero point (Stetson
1998), LMC true modulus (Freedman et~al. 2001), and LMC apparent PL
relations (Udalski et~al. 1999) described above, we re-fit PL relations
for each galaxy to the identical Cepheids employed by Gibson et~al.
(2000a).\footnote{This differs slightly from that of Freedman et~al.
(2001) who re-fit PL relations for each galaxy, but employed Cepheid
samples which differed from those published in the earlier papers in the
series.  This difference is inconsequential though in terms of its effect
upon H$_0$.} Table~2 shows the results of our re-fitting procedure
assuming no-metallicity dependence in the Cepheid PL relation
($\mu_\circ$) and assuming a mild metallicity dependency of $\gamma_{\rm
VI}$=$-$0.2\,mag\,dex$^{-1}$ ($\mu_{\rm Z}$).  The number of Cepheids
employed in the fit n$_{\rm Ceph}$ and the internal random uncertainties
$\sigma_\mu$(r) are also listed.

The primary source of the difference between the distances listed here in
Table~2, versus those presented earlier in Gibson et~al. (2000a), comes
from the adoption of the Udalski et~al. (1999) LMC PL relations (in lieu
of those of Madore \& Freedman 1991).  The Udalski et~al. relations are
based upon the homogeneous OGLE dataset, while the Madore \& Freedman
relations were derived from a more heterogeneous sample culled from a
variety of sources.  The former also benefit from exceedingly well-sampled
light curves, and a factor of three greater number of Cepheids employed in
the fitting.  The Udalski et~al.  $I$-band PL relation
is $\sim$0.10\,mag\,dex$^{-1}$ flatter in slope which acts
to reduce the distances of the SN calibrator galaxies by
$\sim$8\% in the mean.  The OGLE photometry and re-fitted LMC PL relations
has been confirmed by the independent analysis of Sebo et~al. (2001, in
preparation).

Using the SN photometry and reddenings tabulated by Gibson et~al. (2000a;
Table~5), supplemented now with the apparent peak magnitudes (Lira et~al.
1998; Table~7) and Galactic$+$intrinsic reddenings (Phillips et~al. 1999;
Table~2) for SN~1991T, the $B$-, $V$-, and $I$-band peak luminosities for
the nine calibrating SNe can be calculated.  These are listed in columns
3-5 of Table~3, assuming the metallicity-corrected true moduli $\mu_{\rm
Z}$ listed in Table~2.  While not shown, the corrected peak luminosity for
SN~1991T (e.g. M$_{\rm B,corr}^{\rm max}$=$-$19.40$\pm$0.24) is
indistinguishable from that of the mean of the full sample of nine
calibrators ($<$M$_{\rm B,corr}^{\rm max}$$>$=$-$19.32$\pm$0.08).  While
only one datum, there exists no evidence to suggest that spectroscopically
peculiar Type~Ia SNe need be dismissed \it a priori \rm from future
extragalactic distance scale work.

These calibrated peak luminosities can be used in conjunction with the
Hubble relations described by equation~\ref{eq:eq1} to provide SN and
color-dependent Hubble constants.  Columns 6-8 of Table~3 list
H$_0$(B,V,I) for each of the nine calibrators.  The weighted mean of
H$_0$(B), H$_0$(V), and H$_0$(I), for the full sample, yields
H$_0$=73\,km\,s$^{-1}$\,Mpc$^{-1}$, with a \it random \rm uncertainty of
$\pm$2\,km\,s$^{-1}$\,Mpc$^{-1}$.

After Freedman et~al. (2001; Table~15), seven sources of error were
incorporated into the \it systematic \rm error budget.  Uncertainties in
the LMC zero point, crowding, and large scale bulk flows each enter in at
the $\pm$0.10\,mag level; the metallicity dependency of the Cepheid PL
relation at the $\pm$0.08\,mag level; the WFPC2 zero point uncertainty at
the $\pm$0.07\,mag level; reddening and bias in the Cepheid PL fitting at
the $\pm$0.02\,mag level each.  In quadrature, the overall systematic
error budget amounts to 0.205\,mag, corresponding to 10\% in H$_0$.
Significantly improving the precision to which we can derive H$_0$ via
Cepheid calibration of secondary distance indicators will require a factor
of two reduction in uncertainty in \it each \rm of these five remaining
dominant sources of systematic uncertainty.  Until then, we are limited to
10\% precision.

Of these seven sources of systematic error, two were not included in the
formal acconting in early papers in the HST Key Project series: that due
to crowding and that due to bulk flows.  Bulk flows were explored in a
preliminary sense by Gibson et~al. (2000a), but neglected in the quoted
systematic error budget (as it was in the companion papers published in
the February 1, 2000 issue of The Astrophysical Journal).  More extensive
modeling, including a more sophisticated multi-attractor model for the
local universe, allows for a $\pm$5\% systematic uncertainty (Freedman et
al. 2001).

The effects of crowding have been claimed to be the dominant source of
error plaguing the Cepheid-based extragalactic distance scale (Stanek \&
Udalski 1999), a claim that has not been substantiated by either empirical
WFPC2 tests (Gibson, Maloney \& Sakai 2000b) or artificial star tests
(Ferrarese et~al. 2000).  Regardless, a systematic uncertainty of $\pm$5\%
was still deemed plausible by Freedman et~al. (2001), and for consistency
with their analysis, we assume the same.

The final important modification to the systematic error budget used by
Gibson et~al. (2000a) is the adoption of a revised LMC true modulus.  
While Gibson et~al. (2000a) assumed $\mu$(LMC)=18.50$\pm$0.13, based upon
a frequentist and Bayesian analysis of the Gibson (2000) compilation of
published LMC true moduli, Freedman et~al. (2001) demonstrate that a
downward revision to $\mu$(LMC)=18.45$\pm$0.10 is appropriate.  This
decrease has been driven primarily by the introduction of three new LMC
distance indicators, all of which support $\mu$(LMC)$<$18.5:  luminosity
of the red clump, eclipsing binaries, and the indirect constraint provided
by the maser distance to NGC~4258.

The remaining four sources of systematic uncertainty have either been
discussed already, or are negligible in comparison with the others (and
for brevity, not discussed further).  Details pertaining to the latter
terms (reddening constraints provided by NICMOS photometry and bias in the
Cepheid PL fits due to short-end period cutoff) are provided by Freedman
et~al. (2001).

In combination, the above random (r) and systematic (s) error budget yields
a final result for the Hubble Constant of
\begin{equation}
{\rm H}_0 = 73\pm 2\,({\rm r})\pm 7\,({\rm s})\,{\rm km\,s^{-1}\,Mpc^{-1}}.
\label{eq:eq2}
\end{equation}
\noindent
Restricting the analysis to the seven best calibrators (\ie, dropping
SN~1960F and 1974G) has no affect on H$_0$.  Ignoring the metallicity
dependency in the Cepheid PL relation (\ie, using $\gamma_{\rm
VI}$=$+$0.0\,mag\,dex$^{-1}$, as opposed to the $\gamma_{\rm
VI}$=$-$0.2$\pm$0.2\,mag\,dex$^{-1}$ employed here) increases H$_0$ by
3\,km\,s$^{-1}$\,Mpc$^{-1}$.

\section{Summary}
\label{summary}

A Cepheid-based distance to NGC~4527, host to the peculiar Type~Ia
SN~1991T, has been derived using the same software pipeline and unbiased
Cepheid PL analysis employed throughout the HST Key Project on the
Extragalactic Distance Scale series of papers. The corrected peak
luminosity is indistinguishable from that of spectroscopically normal SNe,
demonstrating the robustness of the corrected peak luminosity as a
secondary distance indicator.  SN~1991T is only the fourth calibrator SN
for which accurate $I$-band photometry exists.  NGC~4527, in conjunction
with re-derived distances for eight other Type~Ia SN-host galaxies, and a
full accounting of random and systematic uncertainties, yields a robust
value of the Hubble Constant of
H$_0$=73$\pm$2(r)$\pm$7(s)\,km\,s$^{-1}$\,Mpc$^{-1}$.

%\acknowledgements{}

\newpage

\begin{figure}
\includegraphics{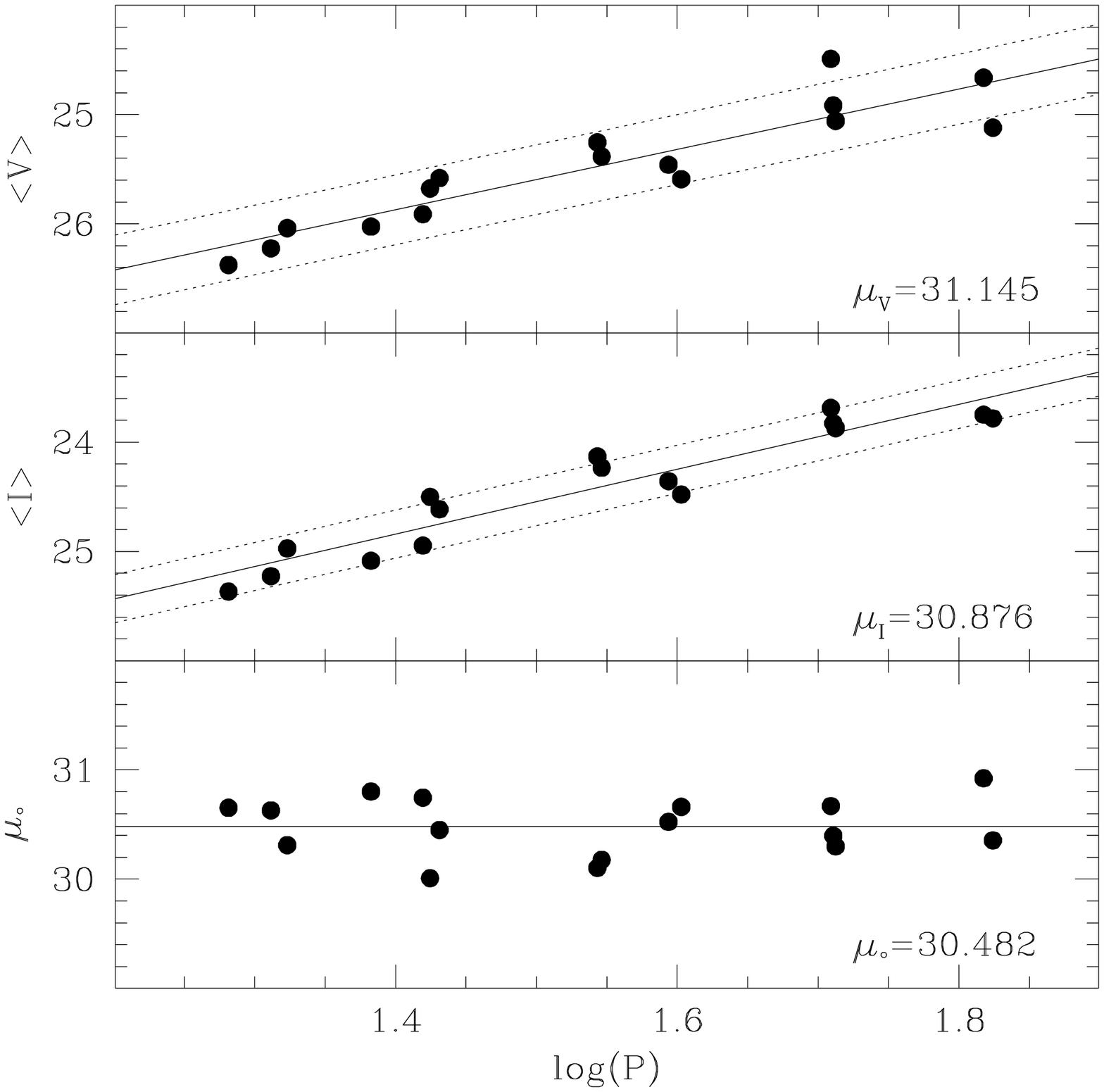}
\caption{Apparent period-luminosity relations in the $V$-
(\it upper panel\rm) and $I$-bands (\it middle panel\rm) based upon the 16
high-quality Cepheid candidates discovered in NGC~4527 (the properties for
which are listed in Table~1).  The solid lines are least-squares fits to
this
entire sample, with the slope fixed to be that of the Udalski et~al. (1999)
LMC
PL relations, while the dotted lines represent their corresponding 2$\sigma$
dispersion.  The inferred apparent distance moduli, ignoring metallicity
effects, are then $\mu_{\rm V}$=31.145$\pm$0.056 (internal) and $\mu_{\rm
I}$=30.876$\pm$0.046 (internal).
\it Lower Panel\rm: Distribution of individually de-reddened Cepheid
true moduli, as a function of period.  The mean corresponds
to $<\mu_\circ>$=30.482$\pm$0.066 (internal).
}
\label{fig:fig1}
\end{figure}

\clearpage

\begin{deluxetable}{ccrrccc}
\tablecaption{Properties of Cepheids Detected in NGC~4527
\label{tbl:tbl1}}
\tablewidth{0in}
\tablehead{
\colhead{ID} & \colhead{Chip} & \colhead{X} & \colhead{Y} &
\colhead{$<$V$>$}  &
\colhead{$<$I$>$}  &
\colhead{$<$P$>$}
}
\startdata
C01 & 1 & 241.4 & 235.9 & 
		       25.38$\pm$0.02 & 24.23$\pm$0.03 & 35.19$\pm$0.32 \nl
C02 & 1 & 375.9 & 236.7 & 
		       26.22$\pm$0.03 & 25.23$\pm$0.06 & 20.49$\pm$0.33 \nl
C03 & 1 & 492.7 & 679.3 & 
		       26.04$\pm$0.03 & 24.97$\pm$0.05 & 21.04$\pm$0.44 \nl
C04 & 1 & 403.6 & 522.0 & 
		       25.68$\pm$0.02 & 24.50$\pm$0.03 & 26.58$\pm$0.35 \nl
C05 & 1 & 485.1 & 489.9 & 
		       25.06$\pm$0.02 & 23.87$\pm$0.03 & 51.58$\pm$1.31 \nl
C06 & 2 &  83.1 & 260.1 & 
		       25.46$\pm$0.02 & 24.36$\pm$0.04 & 39.25$\pm$0.33 \nl
C07 & 2 & 128.2 & 276.2 & 
		       25.91$\pm$0.03 & 24.95$\pm$0.04 & 26.27$\pm$0.42 \nl
C08 & 2 & 138.9 & 251.6 & 
		       24.49$\pm$0.02 & 23.69$\pm$0.02 & 51.17$\pm$0.76 \nl
C09 & 3 & 130.0 & 528.4 & 
		       26.38$\pm$0.03 & 25.37$\pm$0.06 & 19.12$\pm$0.34 \nl
C10 & 3 & 134.6 & 387.5 & 
		       25.59$\pm$0.02 & 24.48$\pm$0.03 & 40.07$\pm$0.63 \nl
C11 & 3 & 369.9 & 265.0 & 
		       24.92$\pm$0.02 & 23.83$\pm$0.03 & 51.38$\pm$2.14 \nl
C12 & 3 & 782.6 & 326.6 & 
		       26.03$\pm$0.03 & 25.08$\pm$0.04 & 24.13$\pm$0.17 \nl
C13 & 4 & 508.9 & 169.4 & 
		       25.12$\pm$0.02 & 23.78$\pm$0.02 & 66.70$\pm$1.43 \nl
C14 & 4 & 391.8 & 670.0 & 
		       25.25$\pm$0.03 & 24.13$\pm$0.04 & 34.94$\pm$0.53 \nl
C15 & 4 & 457.3 & 757.4 & 
		       25.58$\pm$0.03 & 24.61$\pm$0.04 & 26.99$\pm$0.47 \nl
C16 & 4 & 137.3 & 176.4 & 
		       24.66$\pm$0.03 & 23.75$\pm$0.03 & 65.69$\pm$3.24 \nl
\enddata
%\tablenotetext{a}{}
\end{deluxetable}

\clearpage

\begin{deluxetable}{lcccc}
\tablecaption{Adopted True Distance Moduli $\mu_\circ$, 
Metallicity-Corrected Moduli $\mu_{\rm Z}$, and
Internal Random Uncertainties $\sigma_\mu$(r)\tablenotemark{a}
\label{tbl:tbl2}}
\tablewidth{0in}
\tablehead{
\colhead{$\qquad$Galaxy$\qquad$} & 
\colhead{$\qquad$n$_{\rm Ceph}$$\qquad$} &
\colhead{$\qquad\mu_\circ\qquad$} &
\colhead{$\qquad\mu_{\rm Z}\qquad$} &
\colhead{$\qquad\sigma_\mu$(r)$\qquad$}
}
\startdata
NGC~4527  & 16      & 30.482 & 30.562 & $\pm$0.085 \\
NGC~4639  & 17      & 31.524 & 31.624 & $\pm$0.084 \\
NGC~4536  & 27      & 30.693 & 30.763 & $\pm$0.069 \\
NGC~3627  & 17      & 29.794 & 29.944 & $\pm$0.169 \\
NGC~3368  & $\;\,$7 & 29.956 & 30.096 & $\pm$0.098 \\
NGC~5253  & $\;\,$7 & 27.485 & 27.415 & $\pm$0.119 \\
IC~4182   & 28      & 28.207 & 28.187 & $\pm$0.076 \\
NGC~4496A & 51      & 30.750 & 30.804 & $\pm$0.067 \\
NGC~4414  & $\;\,$9 & 31.154 & 31.294 & $\pm$0.105 \\
\enddata
\tablenotetext{a}{All distances are on the Stetson (1998) WFPC2 photometric
zero point, assume the Udalski \etal\ (1999) LMC PL slopes and apparent zero
points, and are referenced to an LMC distance of 49.0\,kpc.  The Cepheids 
employed in the PL fitting are identical to those used by Gibson \etal\ 
(2000a).  Metallicity-corrected distance moduli assume $\gamma_{\rm
VI}=-0.2$\,mag\,dex$^{-1}$ and Cepheid field metallicities as tabulated by
Gibson et~al. (2000a; Table~1).  The random uncertainties are derived
following Gibson et~al. (2000a) and correspond to item $R_{\rm PL}$ of
Table~7
therein.}
\end{deluxetable}

\clearpage

\begin{deluxetable}{lccccccc}
\rotate
\tabletypesize{\footnotesize}
\tablecaption{Values of the Hubble Constant H$_0$
\label{tbl:tbl3}}
\tablewidth{0in}
\tablehead{
\colhead{Galaxy} & 
\colhead{SN} & 
\colhead{M$_{\rm B}^{\rm max}$} &
\colhead{M$_{\rm V}^{\rm max}$} &
\colhead{M$_{\rm I}^{\rm max}$} &
\colhead{H$_0$(B)} &
\colhead{H$_0$(V)} &
\colhead{H$_0$(I)}
}
\startdata
\multicolumn{8}{c}{\sl Metallicity-Corrected: \rm $\gamma_{\rm
VI}=-$0.2\,mag\,dex$^{-1}$} \\
NGC~4527  & 1991T  & $-$19.56$\pm$0.23 & $-$19.59$\pm$0.19 &
$-$19.21$\pm$0.13
		   & 71.7$\pm$9.5      & 68.7$\pm$7.7      & 68.8$\pm$6.9
\\
NGC~4639  & 1990N  & $-$19.36$\pm$0.16 & $-$19.31$\pm$0.13 &
$-$18.91$\pm$0.10
		   & 73.6$\pm$7.8      & 73.7$\pm$6.9      & 74.6$\pm$6.9
\\
NGC~4536  & 1981B  & $-$19.28$\pm$0.15 & $-$19.26$\pm$0.12 &       n/a

		   & 75.1$\pm$8.0      & 74.3$\pm$7.0      &       n/a
\\
NGC~3627  & 1989B  & $-$19.20$\pm$0.25 & $-$19.15$\pm$0.22 &       n/a

		   & 69.2$\pm$9.8      & 69.9$\pm$8.9      &       n/a
\\
NGC~3368  & $\;\,$1998bu & $-$19.42$\pm$0.16 & $-$19.39$\pm$0.14 &
$-$19.12$\pm$0.12
		   & 72.9$\pm$7.9      & 72.2$\pm$7.0      & 68.8$\pm$6.8
\\
NGC~5253  & 1972E  & $-$19.21$\pm$0.22 & $-$19.14$\pm$0.22 &
$-$18.74$\pm$0.23
		   & 87.7$\pm$12.1     & 87.4$\pm$11.4     & 88.4$\pm$12.4
\\
IC~4182   & 1937C  & $-$19.57$\pm$0.28 & $-$19.51$\pm$0.24 &       n/a

		   & 74.4$\pm$11.9     & 74.1$\pm$10.3     &       n/a
\\
NGC~4496A & 1960F  & $-$19.21$\pm$0.32 & $-$19.43$\pm$0.30 &       n/a

		   & 79.8$\pm$13.5     & 70.3$\pm$11.2     &       n/a
\\
NGC~4414  & 1974G  & $-$19.50$\pm$0.32 & $-$19.52$\pm$0.26 &       n/a

		   & 67.5$\pm$11.4     & 65.3$\pm$9.2      &       n/a
\\
\multicolumn{2}{l}{Weighted Mean (All nine SNe)} &                   &
						                     &
						                     &

						   74.0$\pm$3.2      &
						   72.4$\pm$2.8      &
						   72.4$\pm$3.8
\\
\multicolumn{2}{l}{Weighted Mean (Excluding SN~1960F and 1974G)}     & 
						                     &
						                     &
						                     &

						   74.2$\pm$3.5      &
						   73.3$\pm$3.0      &
						   72.4$\pm$3.8
\\
\enddata
\end{deluxetable}

\end{document}